\documentclass[a4paper,fleqn,usenatbib]{mnras}

\usepackage[T1]{fontenc}
\usepackage{ae,aecompl}

\usepackage{graphicx}	
\usepackage{amsmath}	
\usepackage{amssymb}	

\def\msun{\hbox{M$_\odot$}}
\title[eMSTO open clusters]{The extended Main Sequence Turnoff of the Milky Way open cluster 
Collinder\,347}

\author[A.E. Piatti \& Charles Bonatto]{
Andr\'es E. Piatti$^{1,2}$\thanks{E-mail: andres.piatti@unc.edu.ar} and Charles Bonatto$^{3}$\\
$^{1}$Consejo Nacional de Investigaciones Cient\'{\i}ficas y T\'ecnicas, Godoy Cruz 2290, C1425FQB,  Buenos Aires, Argentina\\
$^{2}$Observatorio Astron\'omico de C\'ordoba, Laprida 854, 5000, 
C\'ordoba, Argentina\\
$^{3}$ Departamento de Astronomia,
Universidade Federal do Rio Grande do Sul, Av. Bento Gon\c{c}alves 9500
Porto Alegre 91501-970, RS, Brazil\\
}

\date{Accepted XXX. Received YYY; in original form ZZZ}

\pubyear{2019}

\begin{document}
\label{firstpage}
\pagerange{\pageref{firstpage}--\pageref{lastpage}}
\maketitle

\begin{abstract}
We made use of the {\it Gaia} DR2 archive to comprehensively study the Milky Way open cluster
Collinder\,347, known until now as a very young object of solar metal-content.
However, the $G$ versus $G_{BP}-G_{RP}$ colour-magnitude diagram (CMD) of {\it bonafide} 
probable cluster members, selected on the basis of individual stellar proper motions, their spatial 
distribution and placement in the CMD, reveals the existence of a Hyades-like age open cluster
(log($t$ /yr) = 8.8) of moderately metal-poor chemical content ([Fe/H] = -0.4 dex), with
a present-day mass of 3.3$\times$10$^3$ $\msun$. The cluster exhibits an extended Main Sequence
turnoff (eMSTO) of nearly  500 Myr, while that computed assuming Gaussian distributions
from photometric errors, stellar binarity, rotation and metallicity spread 
yields an eMSTO of $\sim$ 340 Myr.  Such an age difference points
to the existence within the cluster of stellar populations with different
ages.
 \end{abstract} 

\begin{keywords}
(Galaxy:) open clusters and associations: general -- 
(Galaxy:) open clusters and associations: individual -- technique: photometric.
\end{keywords}



\section{Introduction}

Extended Main-Sequence Turnoffs (eMSTOs) have recently been confirmed to be a
common feature in the colour-magnitude diagrams (CMDs)  of Milky Way open clusters 
\citep{cordonietal2018}. It has gained an increasing consensus that its origin could
be linked to stellar rotation \citep{basianetal2018,sunetal2019,lietal2019}.
 Indeed, \citet{marinoetal2018,marinoetal2018b} spectroscopically measured different
rotational velocities of Main Sequence stars of Milky Way open clusters and Magellanic 
Cloud clusters with observed eMSTOs. However, other previous works have concluded that 
age variation, together with rotation, 
are necessary to explain the eMSTO \citep[][and references therein]{gossageetal2019}.
 \citet{dantonaetal2017} have suggested that if the bluest stars 
in the eMSTO are 
interpreted as stars initially rapidly rotating, but that have later slowed down, the age
difference disappears, and "braking" also helps to explain the apparent age differences of
the eMSTO \citep[see also][]{georgyetal2019}.

 Besides eMSTOs, clusters younger than $\sim$ 1 Gyr also exhibit split Main Sequences
\citep[see][and references therein]{miloneetal2018}.
The comparison between the observations and isochrones suggests that the blue Main Sequences
are made of slow-rotating stars, while the red ones host stars with rotational velocities close
to the breakup value. Note that there would be a minimum mass for rotating  stars in order 
to make MSTOs  significantly wider than the rest of the cluster's Main Sequence 
\citep{goudfrooijetal2018}.  Large populations of Be stars, which are fast-rotating stars, 
have also been detected \citep{bastianetal2017,dupreeetal2017}.

Here we report a serendipitous discovery for an open cluster, Collinder\,347, to
exhibit an eMSTO much wider than that predicted considering altogether 
photometric uncertainties, binarity, metallicity spread and stellar rotation. The
finding is twofold, because the cluster was known to be young ($<$ 15 Myr) and
of solar metal content \citep{bukowieckietal2011,kharchenkoetal2013,clariaetal2019},
but we show that it would seem to be a Hyades-like age cluster with a moderately
metal-pool overall abundance. The anomalous wide eMSTO of Collinder\,347
deserves further investigation to answer the  very question about whether it could come
from a intrinsic age spread. If such a hypothesis were confirmed, it would impact
in the way we understand how star clusters could form and evolve.

In Section 2 we describe the used data sets and the applied filtering criteria in order to
select probable cluster members. Section 3 deals with the cluster structural and
astrophysical properties estimates, while in Section 4 we discuss the unveiled cluster
eMSTO. 

\section{Data handing}

\begin{figure*}
\includegraphics[width=\textwidth]{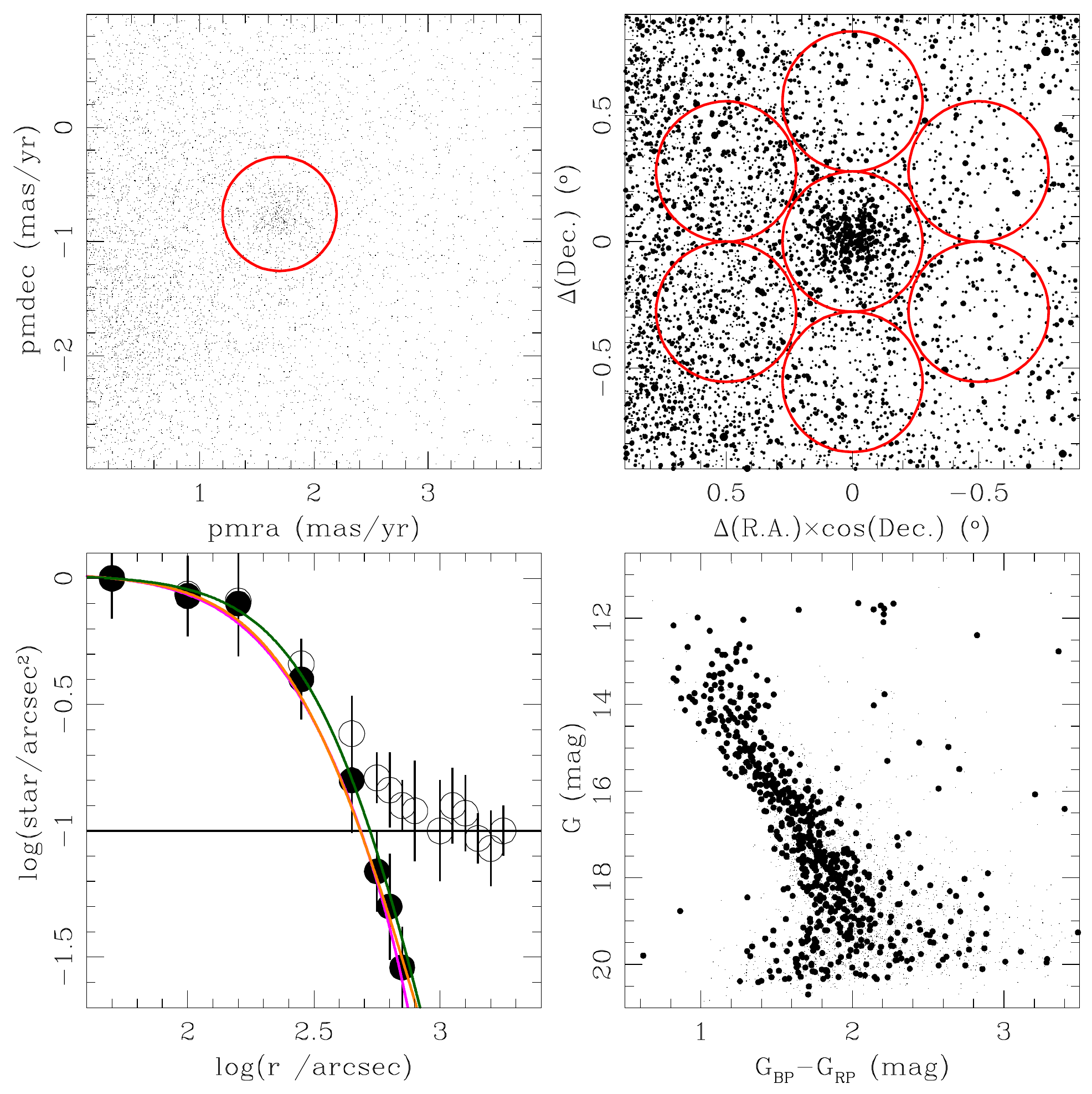}
\caption{{\it Top-left panel:} Vector {\it Gaia} DR2's proper motions diagram of stars in the field of 
Collinder\,347. The red circle's radius is 0.5 mas/yr. {\it Top-right panel:} Spatial distribution
of stars located within the drawn circle in the vector proper motion diagram. The size of the
symbols is proportional to the $G$ mag. The red circles' radii are 0.28$\deg$
(log(r /arcsec) = 3). {\it Bottom-left panel:} Normalised observed and background
subtracted stellar  density radial profiles drawn with open and filled symbols,
respectively. The horizontal line corresponds to the adopted mean background level, 
while the magenta, orange and green lines are the best-fitted \citet{king62},
\citet{plummer11} and \citet{eff87} models, respectively. {\it Bottom-right:} cluster
CMD for stars located within both, the circle in the vector proper motion diagram and
the central cone in the finding chart drawn with large filled symbols; the small ones
correspond to stars distributed within a radius of 0.84$\degr$ from the cluster centre.}
\label{fig:fig1}
\end{figure*}

We used the {\it Gaia} DR2 archive\footnote{http://gea.esac.esa.int/archive/} to retrieve astrometric
positions (RA,DEC,{\it l},b), proper motions in Right Ascension (pmra) and Declination (pmdec)
and $G,G_{BP},G_{RP}$ photometry of stars in a field of 2$\degr$$\times$2$\degr$ centred on
Collinder\,347. We limited our sample to stars with excess noise (\texttt{epsi} $<$ 2) and 
significance of excess of noise (\texttt{sepsi} $<$ 1) to prune the data 
\citep{lindegrenetal2018,piattietal2019}. Top-left panel of Fig.~\ref{fig:fig1} shows a
group of stars inside the drawn circle that arises as an overdensity in the vector
proper motion diagram; they clearly reveal Collinder\,347 on the sky (see the top-right panel). Furthermore,
the group of stars encircled within the central circle  mainly define the cluster CMD
(bottom-right panel), which seems to correspond to a populous and intermediate-age
open cluster. The radius of that circle is the cluster radius (log($r_{cls}$ /arcsec) = 3.0), defined
as the intersection of the cluster stellar radial profile with the mean background level
(see bottom-left panel of Fig.~\ref{fig:fig1}).

\begin{figure*}
\includegraphics[width=\columnwidth]{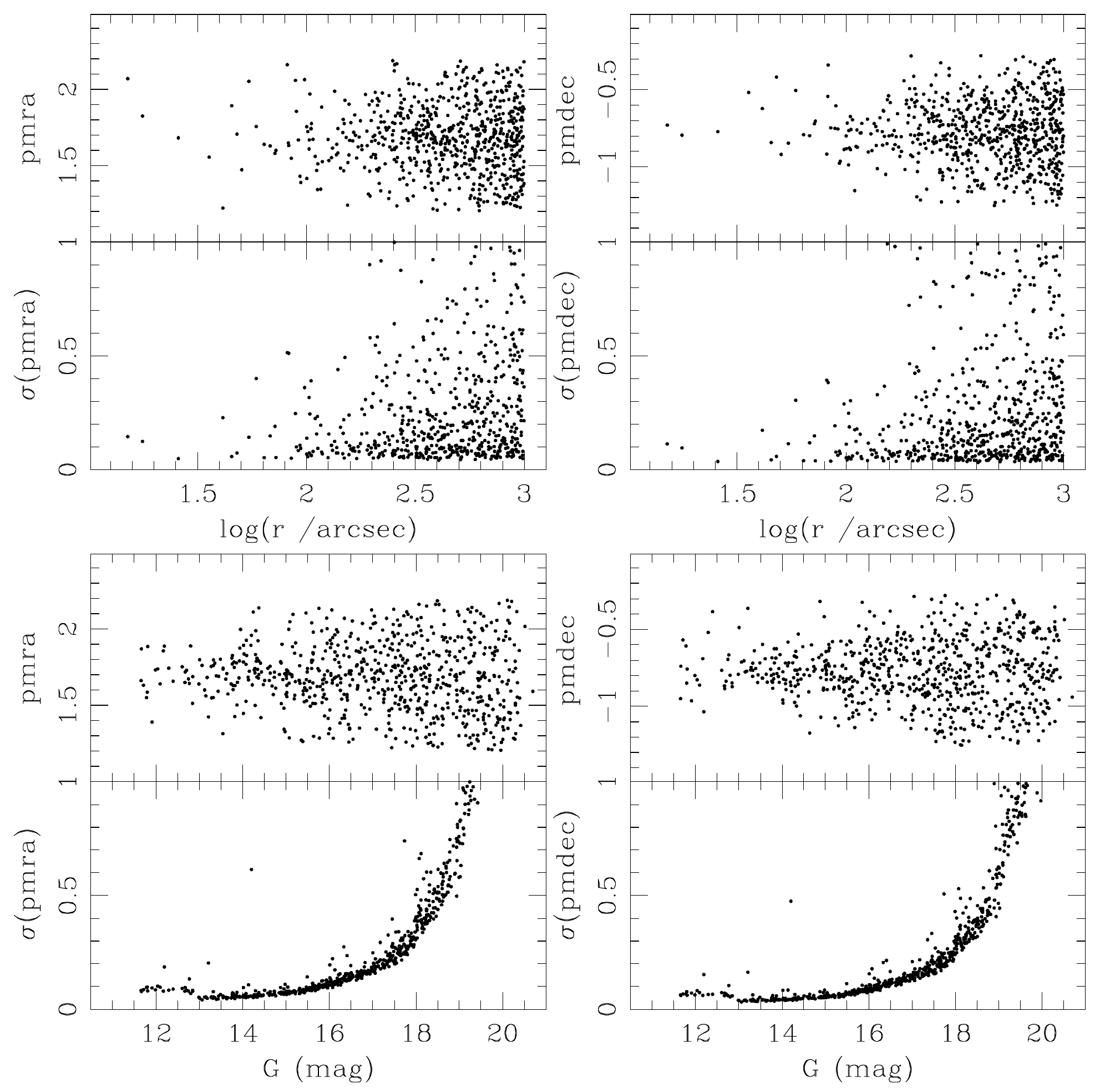}
\includegraphics[width=\columnwidth]{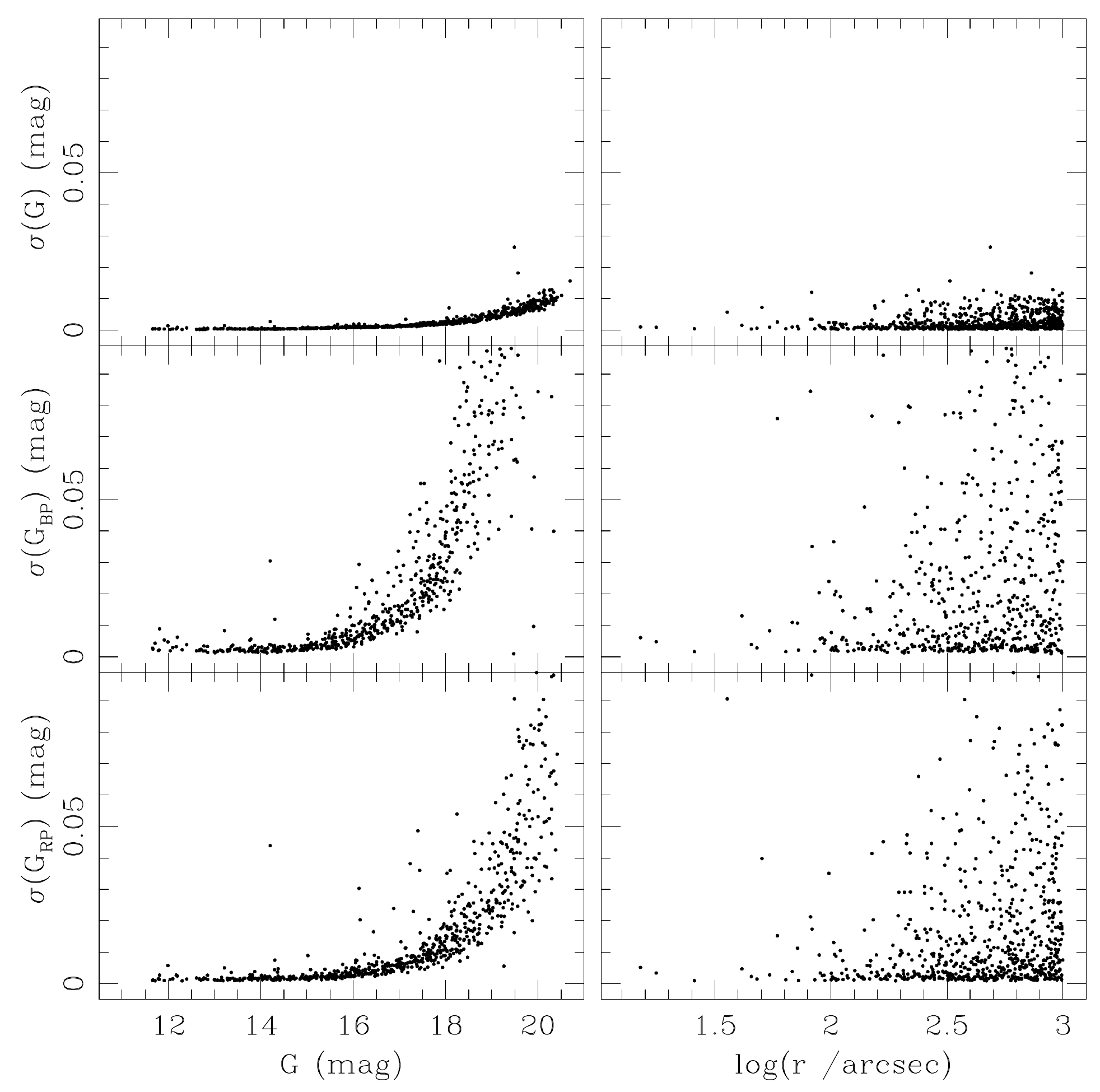}
\caption{Kinematic and photometric data and their respective uncertainties as a
function of the star brightness and distance from the cluster centre for stars
located within both, the circle in the vector proper motion diagram and
the central one in the finding chart (see Fig.~\ref{fig:fig1}).}
\label{fig:fig2}
\end{figure*}

We first inspected to quality of the kinematic and photometric information of all the stars
located inside both, the circle in the vector proper motion diagram and the central circle in 
the finding chart. Fig.~\ref{fig:fig2} shows that there is no dependence of the individual
proper motions with the $G$ magnitude nor with the distance $r$ from the cluster
centre.  Their uncertainties increase with the $G$ magnitude, as expected.  As for the
{\it Gaia} DR2 photometry, there would seem to be no trend with crowding. Furthermore,
according to \citet[][see Section 3]{arenouetal2018}, the {\it Gaia} DR2 photometry
completeness is $\ga$ 90 per cent for stars with $G$ $<$ 19 mag in the inner
region of a globular cluster with $\sim$ 10$^4$ stars/sq deg, so that we deal with
basically a complete photometry data sets.

In order to clean the cluster CMD from the contamination of field stars with
similar cluster kinematics projected along the cluster line-of-sight, we used
as reference six circular regions distributed around the cluster circle as shown in
Fig.~\ref{fig:fig1} (top-right panel). For each reference star field CMD, we 
generated a sample of boxes ($G_o$, ($G_{BP}$ - $G_{RP}$)$_o$) centred 
on each star, with sizes ($\Delta$($G$), $\Delta$($G_{BP}$ - $G_{RP}$))
defined in such a way that one of their corners coincides with the
closest star in that CMD region. The intrinsic $G_o$ magnitudes and
($G_{BP}$ - $G_{RP}$)$_o$ colours where computed from individual
$E(B-V)$ reddenings obtained from the interstellar absorption maps
produced by \citet{lallementetal2019}, and the relationships 
$A_G$ = 2.44 $E(B-V)$ and $E(G_{BP} - G_{RP})$= 1.27 $E(B-V)$
\citep{cetal89,wch2019}.  \citet{lallementetal2019} used 
the {\it Gaia} DR2 and 2MASS photometric data sets to estimate
the extinction towards $\sim$27 millions carefully selected
stars, from which a 3D map of Milky Way interstellar dust within
3 kpc from the Sun was generated. The mean differential-reddening
resolution is 0.003 mag/pc.

Details of the procedure of representing 
the reference star field CMD with an assembly of boxes can be found in
\citet{pb12} and \citet{petal2018}. It has the advantage
of accurately reproducing the reference star field in terms of stellar
density, luminosity function and colour distribution.  Note that
this also implies to consider the variations of the interstellar reddening
across the cluster field.
The generated box sample of each reference star field CMD was superimposed at a time
to the cluster CMD and subtracted from it one star per box; that
closest to the box centre. We merged all the six cleaned cluster CMDs and
produced one with membership probabilities ($P$) on the basis of the number of times
a star kept unsubstracted.  In the subsequent analysis we only retain
star with $P >$ 90 per cent, i,e., the most probable cluster members
as judged by their kinematics, spatial distribution in the cluster field
and position in the cluster CMD.

\begin{figure*}
\includegraphics[width=\textwidth]{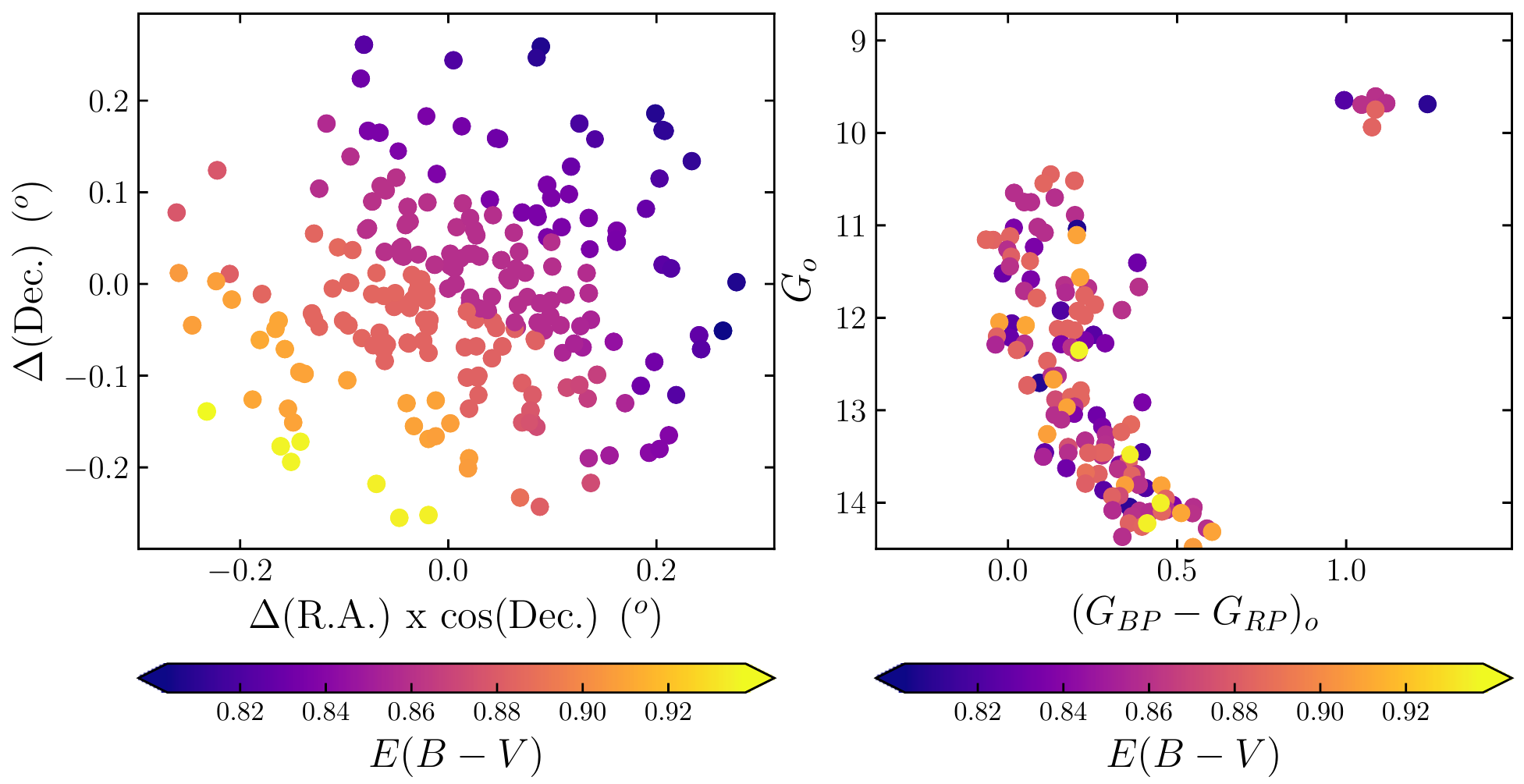}
\caption{Spatial distribution (left panel) and cluster CMD (right panel)
for stars with $P >$ 90 per cent, on which our analysis relies. Colour-coded
symbols represent the individual $E(B-V)$ values, while their sizes are
proportional to $\sigma$($E(B-V)$) in the range  0.095-0.0115 mag.}
\label{fig:fig3}
\end{figure*}

Fig.~\ref{fig:fig3} illustrates the spatial distributions (left panel) and the
intrinsic cluster CMD (right panel) of stars with $P >$ 90 per cent, respectively.
We have coloured the stars according to the reddening values provided by \citet{lallementetal2019}, which were used to obtain the intrinsic magnitudes
and colours. 
As can be seen, the total differential reddening across the cluster field 
amounts $\sim$ 0.14 mag, which represents a spatial reddening variation of
$\sim$ 0.003 mag/pc (= 0.006 mag/arcmin). The errors of the individual $E(B-V)$ values
spans the range 0.095 - 0.115 mag, which could blur any signature
in the cluster CMD with a resolution smaller than $\sim$ 0.10 mag. Nevertheless, it does not appear to be any differential
reddening effects affecting the distribution of stars along the cluster's Main Sequence. For instance, reddest Main Sequence stars -- particularly
those grouped around $G_o$ $\approx$ 12.0 mag and 
$(G_{BP} - G_{RP})_o$ $\sim$ 0.25 mag -- are spread over the whole cluster field, i.e., they come from cluster regions with different $E(B-V)$ values.

\section{cluster's astrophysical properties}

\begin{figure*}
\includegraphics[width=\textwidth]{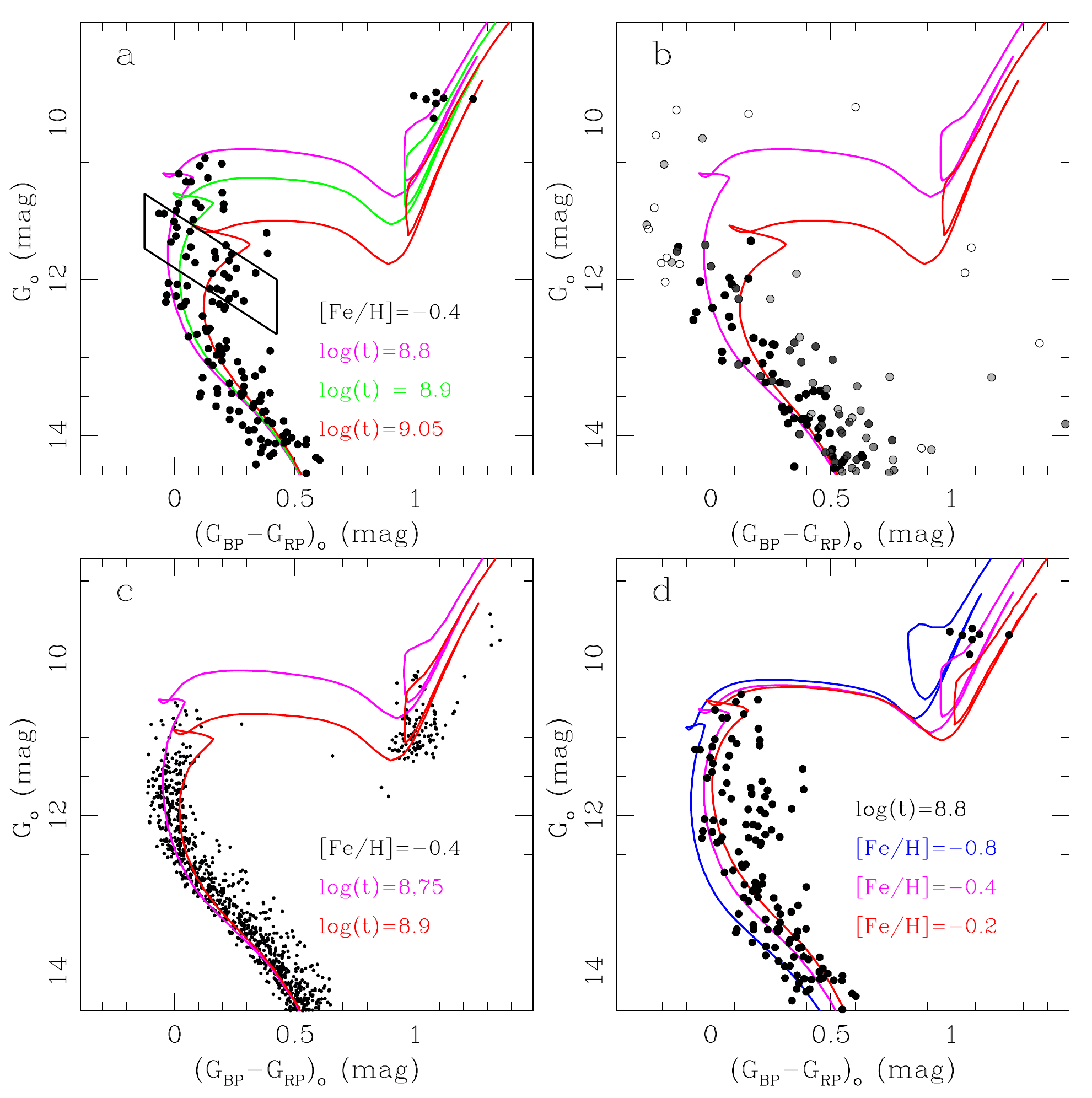}
\caption{Cleaned intrinsic cluster CMD for stars with $P >$ 90 per cent (panel a) and
stars with $P <$ 90 per cent (panel b).  A rectangle scanning the eMSTO region is also superimposed (panel a). Both panels show the same isochrones
superimposed. A synthetic CMD is shown in panel c, while panel d is the same
as panel a with different set of isochrones overplotted (see text for details).}
\label{fig:fig4}
\end{figure*}

Fig.~\ref{fig:fig4} (panel a) shows the resulting intrinsic cluster CMD for stars
with $P >$ 90 per cent. At first glance, there is no evidence for Collinder\,347 
to be a young open cluster, but  one of intermediate-age.  In order to 
estimate its age, distance and metallicity, we used that CMD to which we
superimposed a subset of theoretical  isochrones computed by \citet{betal12}.
Taking into account  as a reference the colour difference between the red 
clump and the MSTO, and the curvature and its extent of the lower envelope 
of the upper Main Sequence,  we restricted the isochrone sample to 
those with 8.5 $<$ log($t$ /yr)  $<$ 9.1, in steps of 
$\Delta$(log($t$ /yr)) = 0.05, and metallicities spanning almost all the
Galactic open cluster metallicity range, namely, -0.8 $<$ [Fe/H] (dex) $<$ +0.2
\citep{hetal14}, in steps of $\Delta$[Fe/H] = 0.10 dex. We fitted all possible isochrone combination and
found that it is necessary to allow  isochrones spanning a range of ages
in order to satisfactorily reproduce the upper Main Sequence bluest
and reddest envelopes, simultaneously, provided that all of them are fitted
using the the same true distance modulus and metallicity.
Particularly, we found that  $<(m-M)_o>$ = (10.70 $\pm$ 0.15) mag, equivalent to a Galactocentric distance of $R_{GC}$ = 6.9 kpc, 
$<$[Fe/H]$>$ = (-0.40 $\pm$ 0.15) dex and log($t$ /yr) = 8.80 - 9.05, provide with
the best resemblance of the cluster CMD. The youngest isochrone
reproduces at a $\sigma$(log($t$ /yr)) = 0.05 level the bluest lower and upper parts of the cluster Main Sequence,
simultaneously, while that of log($t$ /yr)= 8.9 could be associated to
the mean fiducial cluster Main Sequence.
The  cluster CMD also hints at the existence of an eMSTO, as judged by the satisfactory
match of  the isochrone of log($t$ /yr) = 9.05. 

\begin{figure}
\includegraphics[width=\columnwidth]{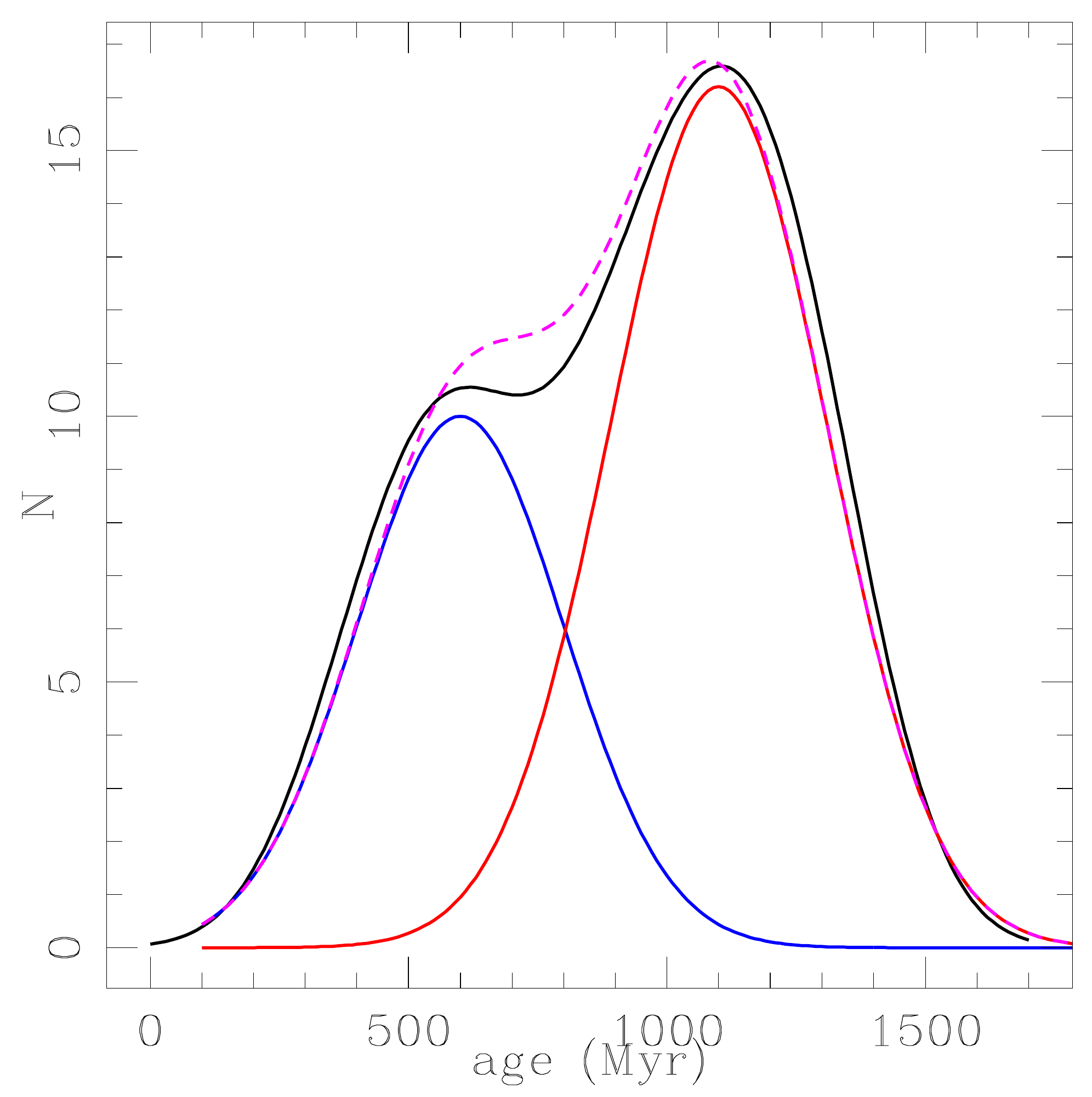}
\caption{Number of stars counted along the major axis of the rectangle
drawn in Fig.~\ref{fig:fig4} shown with a black curve.  Blue and red curves represent the
fitted Gaussians to the observed bimodal distribution, while the dashed magenta curve is the sum of both of them.}
\label{fig:fig5}
\end{figure}

In order to quantify the extent of the eMSTO we 
counted the number of stars ($N$) located within the rectangle drawn in 
Fig.~\ref{fig:fig4} (panel a), using as independent variable the coordinate
(X) along the long axis \citep{goudfrooijetal11, p13}. We then built the $N$ distribution  by summing all
the individual X values. We represented each X value by a Gaussian
function with centre and full-width half maximum equal to the X value and 
2.355 times the associated error $\sigma$(X), respectively, and assigned 
to each Gaussian the same mean intensity. $\sigma$(X) was computed
taking into account the photometric errors (see Fig.~\ref{fig:fig2}) and
the individual $E(B-V)$ errors. Thus, we avoid
the problem of the histogram dependence on the bin size and the end points of
bin, which frequently lead to a difficulty in interpreting the
results. The result is a continuous distribution -- instead of a
discrete histogram -- that allows to appreciate the finest
structures. Fig.~\ref{fig:fig5} depicts the resulting $N$ distribution with a black
curve. As can be seen, a bimodal profile emerges, with blue and red
peaks at 600$\pm$30 Myr and 1100$\pm$30 Myr, respectively, as
fitted by the {\sc iraf.ngaussfit} routine. We
then adopted as a mean eMSTO extent the difference between both mean 
values, i,e. 500$\pm$60  Myr.

For completeness purposes. we exploited the {\it Gaia} DR2 data sets to build the cluster
stellar radial density profile. We used all the stars retrieved in the 2$\degr$$\times$2$\degr$ 
region centred on the cluster with proper motions within the circle drawn in Fig.~\ref{fig:fig1}
(top-left panel) and brighter than $G =$ 18 mag, to secure photometry completeness. We 
then counted the number of stars in rings centred on the cluster of  width 15$\arcsec$ up to 
110$\arcsec$ in steps of 15$\arcsec$, and averaged all the constructed individual radial 
profiles. Fig.~\ref{fig:fig1} (bottom-left panel) depicts the resulting radial profile with
open circles, while the background subtracted one -- the mean background level was estimated from
rings far away the cluster region -- is drawn with filled circles. We fitted \citet{king62},
\citet{plummer11} and \citet{eff87} models by $\chi^2$ minimisation and obtained
core ($r_c$), half-light ($r_h$) and tidal ($r_t$ ) radii of  1.60$\pm$0.15 pc, 2.80$\pm$0.15 pc and
8.00$\pm$1.50 pc, respectively. The EFF's power-law at large radii ($\gamma$) turned out ot be 
5.0$\pm$0.5. 

If the cluster mass were confined to $r_t$, and a Milky  Way mass  enclosed
at the cluster Galactocentric distance were assumed to be 
log($M(< R_{GC})$ = 11.41 + 0.527$\times$log($R_{GC})$
= 11.85 \citep{tayloretal2016}, the present-day cluster's mass from \citet{cw90}
would result (3.3$\pm$1.8)$\times$10$^3$ $\msun$. From \citet[][eqs. 2, 3, 7 and 11]{lamersetal2005a} 
and \citet[][Table 1]{lamersetal2005b}, we found that the fraction of initial mass still bound
to the cluster is $\sim$ 0.04, after lived more than $\sim$ 20 times its mean relaxation time
$t_h$ = 32$\pm$4 Myr, computed from \citet[][eq. 5]{sh71} for a mean
MSTO stellar mass of 2.8 $\msun$.

\section{Analysis and discussion}

The reddest MSTO stars observed in Fig.~\ref{fig:fig4},a unlikely belong to the field, 
since every stars located
within the cluster's circle and with $P <$ 90 per cent have intrinsic magnitudes
and colours clearly different. Fig.~\ref{fig:fig4},b depicts the loci of these
stars with  grey-coded symbols, where black and white filled circles 
corresponding to 85 $<$  $P$ (\%) $<$ 90 and 10 $<$ $P$ (\%) $<$ 15, respectively.
We thoroughly also investigated whether the  photometric errors and the individual stellar 
reddening errors could blur such an eMSTO. To assess on this effect we
performed Monte Carlo simulations to generate a thousand CMDs per
star with $P >$ 90 per cent, allowing that star to have random intrinsic magnitudes
and colours  within the interval [$-\Delta(G),+\Delta(G)$] and 
[$-\Delta(G_{BP}-G_{RP}),+\Delta(G_{BP}-G_{RP})$], respectively. We 
computed: \\

\noindent $\Delta(G)$ = [$\sigma$$_G$$^2$ + (2.44 $\sigma$$_{E(B-V)}$)$^2$]$^{1/2}$, \\

\begin{figure}
\includegraphics[width=\columnwidth]{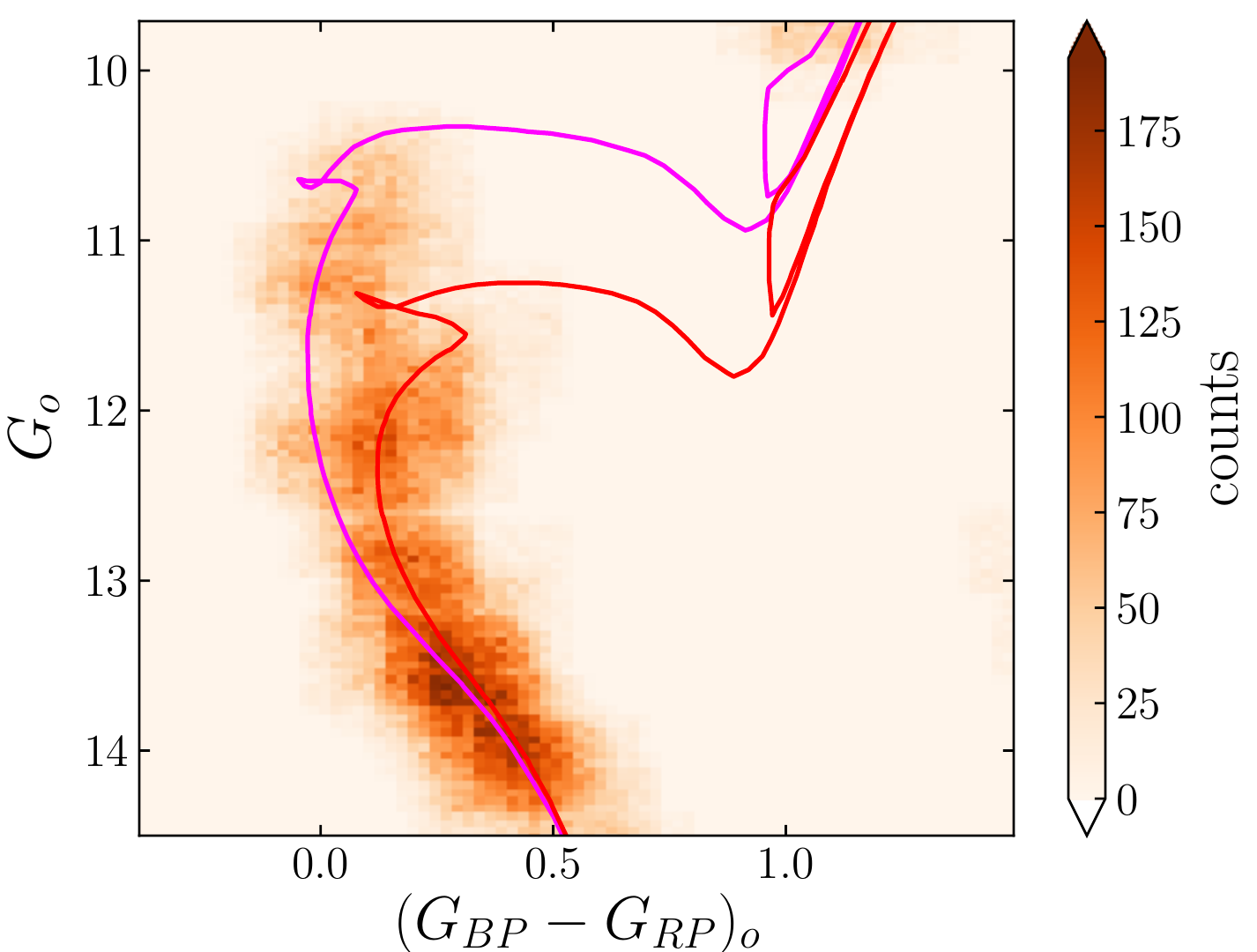}
\caption{Hess diagram of Fig.~\ref{fig:fig4} (panel a) generated considered
photometric errors and individual stellar $E(B-V)$ errors.}
\label{fig:fig6}
\end{figure}

\noindent $\Delta(G_{BP}-G_{RP})$ = [$\sigma$$_{G_{BP}}$$^2$ + 
$\sigma$$_{G_{RP}}$$^2$ + (1.27 $\sigma$$_{E(B-V)}$)$^2$]$^{1/2}$, \\

\noindent where $\sigma$ represents the errors in each quantity.
From a total of 227000 generated CMDs we built the Hess diagram of Fig.~\ref{fig:fig6},
which still exhibits  the eMSTO seen in Fig,~\ref{fig:fig4}.
Note that with more accurate reddening estimates, the eMSTO feature would have
resulted even more visible.

The combined effect of stellar binarity and photometric errors was evaluated through
the generation of a synthetic CMD for a cluster with a total mass of 
5$\times$10$^3$ $\msun$ and age and metal content as for Collinder\,347. 
We used {\sc syclist}\footnote{https://www.unige.ch/sciences/astro/evolution/en/database/syclist/} \citep{ekstrometal12,georgyetal14} and {\sc padova.cmd}\footnote{http://stev.oapd.inaf.it/cgi-bin/cmd} \citep{betal12} packages and assumed
a binary faction of 0.5 \citep{miloneetal16,pc2017}. The resulting synthetic
CMD is shown in Fig.~\ref{fig:fig4},c where we superimposed two
theoretical isochrones \citep{betal12} embracing the observed spread. 
Following the same procedure described above to estimate the
observed MSTO extent, we obtained from the synthetic CMD $\Delta$(age) = 230$\pm$30 Myr.

We
also dealt with metallicity spread using a subset of isochrones, which assume
$Y = 0.2485 + 1.78 Z$ (see Fig.~\ref{fig:fig4},d). Note that we assumed
an arbitrary amount of metallicity and helium content variations, aiming at illustrating
for completeness purposes that they do not impact on the MSTO extent.
Indeed, as far as we are aware, there is no evidence of metallicity and
helium spread in Milky Way open clusters \citep{donoretal2018}. In this
case, we adopted $\Delta$(age) = 50$\pm$10 Myr.

As for estimating the magnitude of stellar rotation effects, we made use of the
relationship obtained by \citet{niederhoferetal15b} between the age of a star
cluster and the expected (apparent) age spread. Thus, by interpolating
an age of log($t$ /yr) = 8.8) in their Fig. 2, we obtained  $\Delta$(age) = 
250$\pm$40 Myr. High fraction of Be stars that implies a high fraction of rapidly rotating stars
have been found in Milky Way open clusters and Magellanic Cloud clusters
\citep{bastianetal2017,miloneetal2018}. As H$\alpha$ contributes to the three
{\it Gaia} DR2 bandpasses\footnote{https://www.cosmos.esa.int/web/gaia/iow\_20180316}, and 
hence it is in principle cancelled through the $G_{BP} - G_{RP}$ colour,  the observed broadness
-- particularly the redest stars around $G_o$ $\approx$ 12.0 mag and $(G_{BP} - G_{RP})_o$ $\sim$
0.25 mag -- confirms that the observed spread is not mainly caused by rapid rotators.

Finally, we added in quadrature all the apparent age spreads by approximating them
to Gaussian distributions \citep[see][]{pc2017}, namely, that from observational
errors and stellar binarity (230 Myr), that from stellar rotation (250 Myr), and that
from iron-helium variations (50 Myr), to obtain an overall apparent age spread of   340$\pm$50 Myr.
As can be seen, the resulting value is clearly smaller than the observed one of 500$\pm$60Myr. The difference between the mean values is 160 Myr,
clearly smaller than the sum of their respective uncertainties (100 Myr).
Moreover, if we subtracted them in quadrature, then the intrinsic age
spread would turn out to be 360$\pm$130 Myr.
This extended observed age spread points to the need of considering  stellar populations with different ages could populate Collinder\,347. 

\section*{Acknowledgements}
We thank Mateus Angelo and Wilton Dias because their work inspired us
to dig into the {\it Gaia}' archive, and Dafydd Evans to provided me with
feedback about {\it Gaia}'s photometry completeness.
This work presents results from the European Space Agency (ESA)
space mission Gaia. Gaia data are being processed by the Gaia Data Processing
and Analysis Consortium (DPAC). Funding for the DPAC is provided by national
institutions, in particular the institutions participating in the Gaia
MultiLateral Agreement (MLA). The Gaia mission website is
\url{https://www.cosmos.esa.int/gaia}. The Gaia archive website is
\url{https://archives.esac.esa.int/gaia}.










\bsp	
\label{lastpage}
\end{document}